\documentclass[%
reprint,
superscriptaddress,
amsmath,amssymb,
aps,
prl
]{revtex4-1}

\usepackage{graphicx}
\usepackage{subfigure}
\usepackage{siunitx}
\usepackage{xcolor}
\usepackage[utf8x]{inputenc}
\usepackage{supertabular}
\usepackage{hyperref}
\usepackage{soul}

\begin{document}

\title{First Search for the Absorption of Fermionic Dark Matter with the PandaX-4T Experiment}


\def\shKeyLab{School of Physics and Astronomy, Shanghai Jiao Tong University, MOE Key Laboratory for Particle Astrophysics and Cosmology, Shanghai Key Laboratory for Particle Physics and Cosmology, Shanghai 200240, China}
\def\BUAA{School of Physics, Beihang University, Beijing 102206, China}
\def\BUAALab{Beijing Key Laboratory of Advanced Nuclear Materials and Physics, Beihang University, Beijing, 102206, China}
\def\zzu{School of Physics and Microelectronics, Zhengzhou University, Zhengzhou, Henan 450001, China}
\def\USTClab{State Key Laboratory of Particle Detection and Electronics, University of Science and Technology of China, Hefei 230026, China}
\def\USTCdep{Department of Modern Physics, University of Science and Technology of China, Hefei 230026, China}
\def\BUAALab{International Research Center for Nuclei and Particles in the Cosmos \& Beijing Key Laboratory of Advanced Nuclear Materials and Physics, Beihang University, Beijing 100191, China}
\def\pku{School of Physics, Peking University, Beijing 100871, China}
\def\YaLongSD{Yalong River Hydropower Development Company, Ltd., 288 Shuanglin Road, Chengdu 610051, China}
\def\IAP{Shanghai Institute of Applied Physics, Chinese Academy of Sciences, 201800 Shanghai, China}
\def\CHEPpku{Center for High Energy Physics, Peking University, Beijing 100871, China}
\def\SDUdep{Research Center for Particle Science and Technology, Institute of Frontier and Interdisciplinary Scienc, Shandong University, Qingdao 266237, Shandong, China}
\def\SDUlab{Key Laboratory of Particle Physics and Particle Irradiation of Ministry of Education, Shandong University, Qingdao 266237, Shandong, China}
\def\UMD{Department of Physics, University of Maryland, College Park, Maryland 20742, USA}
\def\TDLee{Tsung-Dao Lee Institute, Shanghai Jiao Tong University, Shanghai, 200240, China}
\def\MESJTU{School of Mechanical Engineering, Shanghai Jiao Tong University, Shanghai 200240, China}
\def\SYU{School of Physics, Sun Yat-Sen University, Guangzhou 510275, China}
\def\SYUSFI{Sino-French Institute of Nuclear Engineering and Technology, Sun Yat-Sen University, Zhuhai, 519082, China}
\def\NKU{School of Physics, Nankai University, Tianjin 300071, China}
\def\FDU{Key Laboratory of Nuclear Physics and Ion-beam Application (MOE), Institute of Modern Physics, Fudan University, Shanghai 200433, China}
\def\USST{School of Medical Instrument and Food Engineering, University of Shanghai for Science and Technology, Shanghai 200093, China}
\def\SJTUSC{Shanghai Jiao Tong University Sichuan Research Institute, Chengdu 610213, China}
\def\Princeton{Physics Department, Princeton University, Princeton, NJ 08544, USA}
\def\MIT{Department of Physics, Massachusetts Institute of Technology, Cambridge, MA 02139, USA}
\def\SARI{Shanghai Advanced Research Institute, Chinese Academy of Sciences, Shanghai 201210, China}
\def\SPEIT{SJTU Paris Elite Institute of Technology, Shanghai Jiao Tong University, Shanghai, 200240, China}

\author{Linhui Gu}\affiliation{\shKeyLab}
\author{Abdusalam Abdukerim}\affiliation{\shKeyLab}
\author{Zihao Bo}\affiliation{\shKeyLab}
\author{Wei Chen}\affiliation{\shKeyLab}
\author{Xun Chen}\affiliation{\shKeyLab}\affiliation{\SJTUSC}
\author{Yunhua Chen}\affiliation{\YaLongSD}
\author{Chen Cheng}\affiliation{\SYU}
\author{Yunshan Cheng}\affiliation{\SDUdep}\affiliation{\SDUlab}
\author{Zhaokan Cheng}\affiliation{\SYUSFI}
\author{Xiangyi Cui}\affiliation{\TDLee}
\author{Yingjie Fan}\affiliation{\NKU}
\author{Deqing Fang}\affiliation{\FDU}
\author{Changbo Fu}\affiliation{\FDU}
\author{Mengting Fu}\affiliation{\pku}
\author{Lisheng Geng}\affiliation{\BUAA}\affiliation{\BUAALab}\affiliation{\zzu}
\author{Karl Giboni}\affiliation{\shKeyLab}
\author{Xuyuan Guo}\affiliation{\YaLongSD}
\author{Ke Han}\affiliation{\shKeyLab}
\author{Changda He}\affiliation{\shKeyLab}
\author{Jinrong He}\affiliation{\YaLongSD}
\author{Di Huang}\affiliation{\shKeyLab}
\author{Yanlin Huang}\affiliation{\USST}
\author{Zhou Huang}\affiliation{\shKeyLab}
\author{Ruquan Hou}\affiliation{\SJTUSC}
\author{Xiangdong Ji}\affiliation{\UMD}
\author{Yonglin Ju}\affiliation{\MESJTU}
\author{Chenxiang Li}\affiliation{\shKeyLab}
\author{Jiafu Li}\affiliation{\SYU}
\author{Mingchuan Li}\affiliation{\YaLongSD}
\author{Shu Li}\affiliation{\MESJTU}
\author{Shuaijie Li}\affiliation{\TDLee}
\author{Qing Lin}\affiliation{\USTClab}\affiliation{\USTCdep}
\author{Jianglai Liu}\email[Spokesperson: ]{jianglai.liu@sjtu.edu.cn}\affiliation{\shKeyLab}\affiliation{\TDLee}\affiliation{\SJTUSC}
\author{Xiaoying Lu}\affiliation{\SDUdep}\affiliation{\SDUlab}
\author{Lingyin Luo}\affiliation{\pku}
\author{Yunyang Luo}\affiliation{\USTCdep}
\author{Wenbo Ma}\affiliation{\shKeyLab}
\author{Yugang Ma}\affiliation{\FDU}
\author{Yajun Mao}\affiliation{\pku}
\author{Nasir Shaheed}\affiliation{\SDUdep}\affiliation{\SDUlab}
\author{Yue Meng}\affiliation{\shKeyLab}\affiliation{\SJTUSC}
\author{Xuyang Ning}\affiliation{\shKeyLab}
\author{Ningchun Qi}\affiliation{\YaLongSD}
\author{Zhicheng Qian}\affiliation{\shKeyLab}
\author{Xiangxiang Ren}\affiliation{\SDUdep}\affiliation{\SDUlab}
\author{Changsong Shang}\affiliation{\YaLongSD}
\author{Xiaofeng Shang}\affiliation{\shKeyLab}
\author{Guofang Shen}\affiliation{\BUAA}
\author{Lin Si}\affiliation{\shKeyLab}
\author{Wenliang Sun}\affiliation{\YaLongSD}
\author{Andi Tan}\affiliation{\UMD}
\author{Yi Tao}\email[Corresponding author: ]{taoyi92@sjtu.edu.cn}\affiliation{\shKeyLab}\affiliation{\SJTUSC}
\author{Anqing Wang}\affiliation{\SDUdep}\affiliation{\SDUlab}
\author{Meng Wang}\affiliation{\SDUdep}\affiliation{\SDUlab}
\author{Qiuhong Wang}\affiliation{\FDU}
\author{Shaobo Wang}\affiliation{\shKeyLab}\affiliation{\SPEIT}
\author{Siguang Wang}\affiliation{\pku}
\author{Wei Wang}\affiliation{\SYUSFI}\affiliation{\SYU}
\author{Xiuli Wang}\affiliation{\MESJTU}
\author{Zhou Wang}\affiliation{\shKeyLab}\affiliation{\SJTUSC}\affiliation{\TDLee}
\author{Yuehuan Wei}\affiliation{\SYUSFI}
\author{Mengmeng Wu}\affiliation{\SYU}
\author{Weihao Wu}\affiliation{\shKeyLab}
\author{Jingkai Xia}\affiliation{\shKeyLab}
\author{Mengjiao Xiao}\affiliation{\UMD}
\author{Xiang Xiao}\affiliation{\SYU}
\author{Pengwei Xie}\affiliation{\TDLee}
\author{Binbin Yan}\affiliation{\shKeyLab}
\author{Xiyu Yan}\affiliation{\USST}
\author{Jijun Yang}\affiliation{\shKeyLab}
\author{Yong Yang}\affiliation{\shKeyLab}
\author{Chunxu Yu}\affiliation{\NKU}
\author{Jumin Yuan}\affiliation{\SDUdep}\affiliation{\SDUlab}
\author{Ying Yuan}\affiliation{\shKeyLab}
\author{Xinning Zeng}\affiliation{\shKeyLab}
\author{Dan Zhang}\affiliation{\UMD}
\author{Minzhen Zhang}\affiliation{\shKeyLab}
\author{Peng Zhang}\affiliation{\YaLongSD}
\author{Shibo Zhang}\affiliation{\shKeyLab}
\author{Shu Zhang}\affiliation{\SYU}
\author{Tao Zhang}\affiliation{\shKeyLab}
\author{Yuanyuan Zhang}\affiliation{\TDLee}
\author{Li Zhao}\affiliation{\shKeyLab}
\author{Qibin Zheng}\affiliation{\USST}
\author{Jifang Zhou}\affiliation{\YaLongSD}
\author{Ning Zhou}\email[Corresponding author: ]{nzhou@sjtu.edu.cn}\affiliation{\shKeyLab}
\author{Xiaopeng Zhou}\affiliation{\BUAA}
\author{Yong Zhou}\affiliation{\YaLongSD}
\author{Yubo Zhou}\affiliation{\shKeyLab}

\collaboration{PandaX Collaboration}
\noaffiliation

\date{\today}

\begin{abstract}
Compared with the signature of dark matter elastic scattering off nuclei, the absorption of fermionic dark matter by nuclei opens up a new searching channel for light dark matter with a characteristic monoenergetic signal.
In this Letter, we explore the $95.0$-day data from the PandaX-4T commissioning run and report the first dedicated searching results of the fermionic dark matter absorption signal through a neutral current process. No significant signal was found, and the lowest limit on the dark matter-nucleon interaction cross section is set to be $1.5\times10^{-50}$~cm$^2$ for a fermionic dark matter mass of $40$~MeV/$c^2$ with 90\% confidence level.
\end{abstract}


\maketitle



Cosmological and astronomical observations strongly indicate the existence of dark matter (DM), but the nature of DM is still a mystery~\cite{Bertone2005}.
While searching for the popular weakly interacting massive particle (WIMP) is in full swing~\cite{Liu:2017drf,Billard:2021uyg,Zhao2020,Meng2021,Akerib2017,Aprile2018,Agnes2018}, other promising DM candidates have also been put forward and searched for experimentally, especially with mass below 1~GeV/$c^2$ ~\cite{Cheng2021,Zhou2021,Xia2019}. However, due to the detection threshold in direct detection experiments, there are no stringent constraints on sub-GeV light DM scattering with the standard model (SM) particles, and the cross section can still be large~\cite{Liu:2017drf,Billard:2021uyg}. Recently, DM absorption scenarios are proposed and studied, which can generate some novel inelastic signatures in direct detection experiments and enhance the sensitivity of light DM searches~\cite{Pospelov:2008jk,An:2014twa,Hochberg:2016ajh,Hochberg:2016sqx,Bloch:2016sjj,Green:2017ybv,Arvanitaki:2017nhi,Dror2019a,Dror2019b,vonKrosigk:2020udi,Dror:2020czw,Mitridate:2021ctr,Li:2022kca}.

In this Letter, we consider a fermionic DM absorption scenario through a neutral current (NC) process with xenon nuclei,  
\begin{equation}
    \overset{(-)}\chi + {}^{A}\mathrm{Xe} \rightarrow \overset{(-)}\nu + {}^{A}\mathrm{Xe} ,
    \label{eq:neutral-current}
\end{equation}
where 
$A$ denotes the atomic mass number of the xenon isotope,
$\chi$($\bar\chi$) is the DM (anti-)particle, and $\nu$($\bar\nu$) is the SM (anti-)neutrino.
Such a scenario can be described by an ultraviolet (UV) complete model with an additional $U(1)$' symmetry breaking~\cite{Dror2019a,Dror2019b}. In this model, a lepton number charged DM $\chi$ mixes with the approximately massless Dirac neutrino $\nu$ through a scalar field $\phi$ Yukawa interaction, giving the $U(1)$' invariant mass term:
\begin{equation}
    \mathcal{L}_{\text{mass}} \supset m_{\chi}\bar{\chi}\chi + \left(y\phi\bar{\chi} P_{R}\nu + \text{h.c.}\right),
    \label{eq:mass-Lagrangian}
\end{equation}
where $y$ is the Yukawa coupling constant. The $\chi$-$\nu$ mixing term naturally derives a completely massless state, which is identified as the SM neutrino, and a massive state with mass $\sqrt{m_\chi^2 + y^2 \langle\phi\rangle^2}$, after diagonalization. The right-handed component $\chi_R \equiv P_R \chi$ is mixed with the right-handed neutrino $\nu_R$ with a mixing angle $\theta_{R}$,
\begin{equation}
    \sin{\theta_{R}} = \frac{y\langle\phi\rangle}{\sqrt{y^{2}\langle\phi\rangle^{2}+m_{\chi}^{2}}}. \quad 
\end{equation}
Through a heavy $Z^{\prime}$ mediator coupled with quarks and $\chi$, the effective operator at the nucleon scale is~\cite{Dror2019a}
\begin{equation}
    \mathcal{O}_\mathrm{NC} = \frac{1}{\Lambda^{2}}\left(\bar{n} \gamma^{\mu} n + \bar{p} \gamma^{\mu} p\right) \bar{\chi} \gamma_{\mu} P_{R} \nu + \text{h.c.} ,
    \label{eq:operator}
\end{equation}
where the energy scale $1 / \Lambda^{2} \equiv Q_{\chi} g_{\chi}^{2} \sin{\theta_{R}} \cos{\theta_{R}} / m_{Z^{\prime}}^{2}$, with $g_{\chi}$ denoting the $U(1)$' gauge coupling, $m_{Z^{\prime}}$ the mediator mass, and
$Q_{\chi}$ the charge of dark matter under the $U(1)$', yielding the absorption process described above.

Similar to the WIMP spin-independent (SI) elastic scattering model, the absorption rate is coherently enhanced for heavy nuclei by a factor of $\sim A^2$, making xenon a preferable target. For a given DM mass, the nuclear recoil (NR) energy in xenon of the fermionic absorption process could be $\sim 10^6$ times larger than the SI elastic scattering, leading to a possibility of searching for a MeV/$c^2$ scale DM with xenon-based experiments.

\begin{figure}[htbp]
    \centering
    \includegraphics[width=0.5\textwidth]{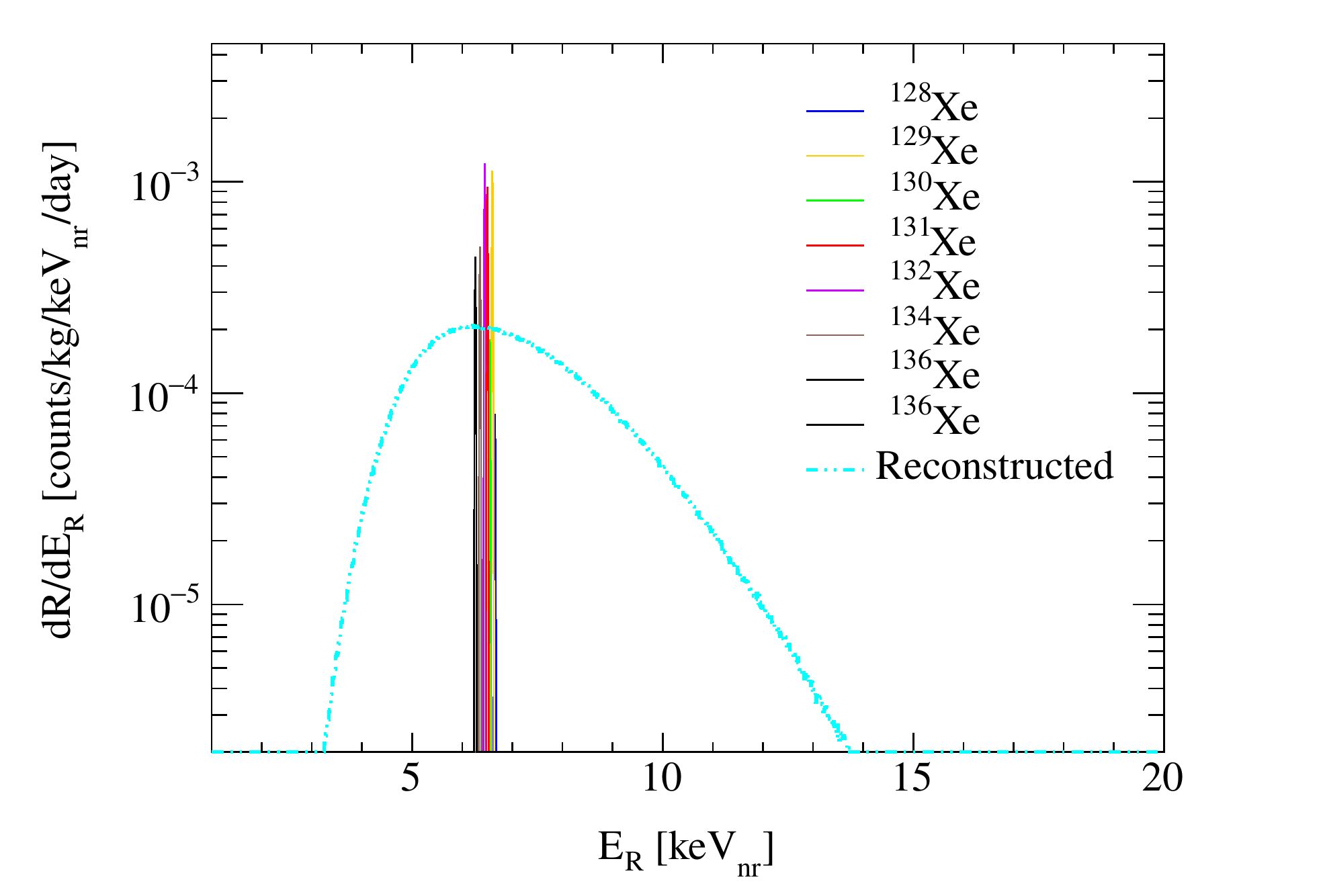}
    \caption{
    Expected fermionic absorption differential rate as a function of NR energy for xenon target, with a DM mass $m_\chi = 40\,\rm MeV$/$c^2$ and absorption cross-section $\sigma_{\chi\mathrm{-N}}^{\mathrm{NC}} = 1.0\times10^{-49}\,\rm cm^2$. Various colors represent contributions from different isotopes of xenon.  The reconstructed energy is shown in cyan, with the Lindhard factor~\cite{lindhard} applied to recover the full NR energy.
    }
    \label{fig:diff-rate}
\end{figure}

With the standard halo model (SHM) for the Earth's nearby DM distribution~\cite{Baxter2021}, the differential event rate of the nuclear absorption signal as a function of recoil energy $E_{R}$  
is given by (with $j$ denoting any specific isotope of xenon)
\begin{equation}
    \frac{d R}{d E_{R}} = \frac{\rho_{\chi}\sigma_{\chi\mathrm{-N}}^{\mathrm{NC}}}{2 m_{\chi}^3 M_{T}} \sum_j \frac{q_{j}}{p_{\nu,j}} N_{j} M_{j} A_{j}^{2} F_{j}^{2}\left\langle\frac{1}{v}\right\rangle_{v>v_{\min,j}},
\end{equation}
where $\rho_{\chi} = 0.3$~GeV/cm$^3$ is the local DM density~\cite{Lewin1996}; $\sigma_{\chi\mathrm{-N}}^{\mathrm{NC}} = m_{\chi}^2 / (4\pi\Lambda^4)$ is the absorption cross section per nucleon; $q_{j} = \sqrt{2 E_{R,j} M_{j}}$ is the momentum transfer to a target nucleus; $p_{\nu,j} = \sqrt{q_{j}(2 m_{\chi} - q_{j} - 2 E_{R,j})}$ denotes the momentum of the outgoing neutrino; $M_{T} = \sum_j N_{j} M_{j}$ is the total target mass with $N_{j}$ and $M_{j}$ corresponding to the number and mass of each isotope, respectively; $A_{j}$ is again the atomic mass number; and $F_{j} \equiv F(q_{j}) $ is the normalized Helm nuclear form factor~\cite{Lewin1996}.
The mass of the nonrelativistic incoming $\chi$ dominates the energy, so the momentum transfer $q_{j} \simeq m_{\chi}$,  
giving a very sharp peak ($E_{R}\simeq m_\chi^2 /2 M_j$) in the NR energy spectrum, with contributions from different xenon isotopes ($j$) slightly offset; see Fig.~\ref{fig:diff-rate}. 

The PandaX-4T experiment, located in the B2 hall at China Jinping Underground Laboratory Phase-II (CJPL-II)~\cite{Kang:2010zza, Li:2014rca}, is a multiphysics purposed xenon experiment~\cite{Zhang2019} aiming to explore DM and neutrino physics. The PandaX-4T detector is a dual-phase xenon time projection chamber (TPC) well shielded by ultrapure water~\cite{Qian2021}, with a sensitive xenon target mass of 3.7~tonne~\cite{Meng2021}. A total of 169 top and 199 bottom 3-inch 
photomultiplier tubes (PMTs) measure the primary prompt scintillation photons ($S1$) and the secondary delayed electroluminescence photons ($S2$) from ionized electrons.
Another two rings of one-inch PMTs are installed outside of the TPC sensitive volume, serving as the veto PMTs for rejecting multiscatter backgrounds.
Unlike an electronic recoil (ER) event, only part of the recoil energy $E_R$ in a NR event is converted into the scintillation photons and ionized electrons, 
which is modeled through the so-called Lindhard factor~\cite{lindhard}. 
The reconstructed energy $E$ from $S1$ and $S2$ of a given event is
\begin{equation}
    E=13.7~{\rm eV} \times (\frac{S1}{\rm PDE}+\frac{S2_{\rm b}}{{\rm EEE} \times {\rm SEG}_{\rm b}}),
    \label{eq:energy}
\end{equation}
in which PDE, EEE, and SEG$_{\rm b}$ are the photon detection efficiency for $S1$, electron extraction efficiency, and the single-electron gain using $S2_{\rm b}$ (the $S2$ collected from the bottom PMT array), respectively, and the 13.7 eV is the mean energy to produce a quanta in liquid xenon. 
The data used in this work consist of five sets during 95.0 calendar days of stable data taking, with different hardware configurations. The parameters used for energy reconstruction in each set are summarized in Ref.~\cite{Meng2021}.

\begin{figure}[htbp]
    \centering
    \includegraphics[width=0.48\textwidth]{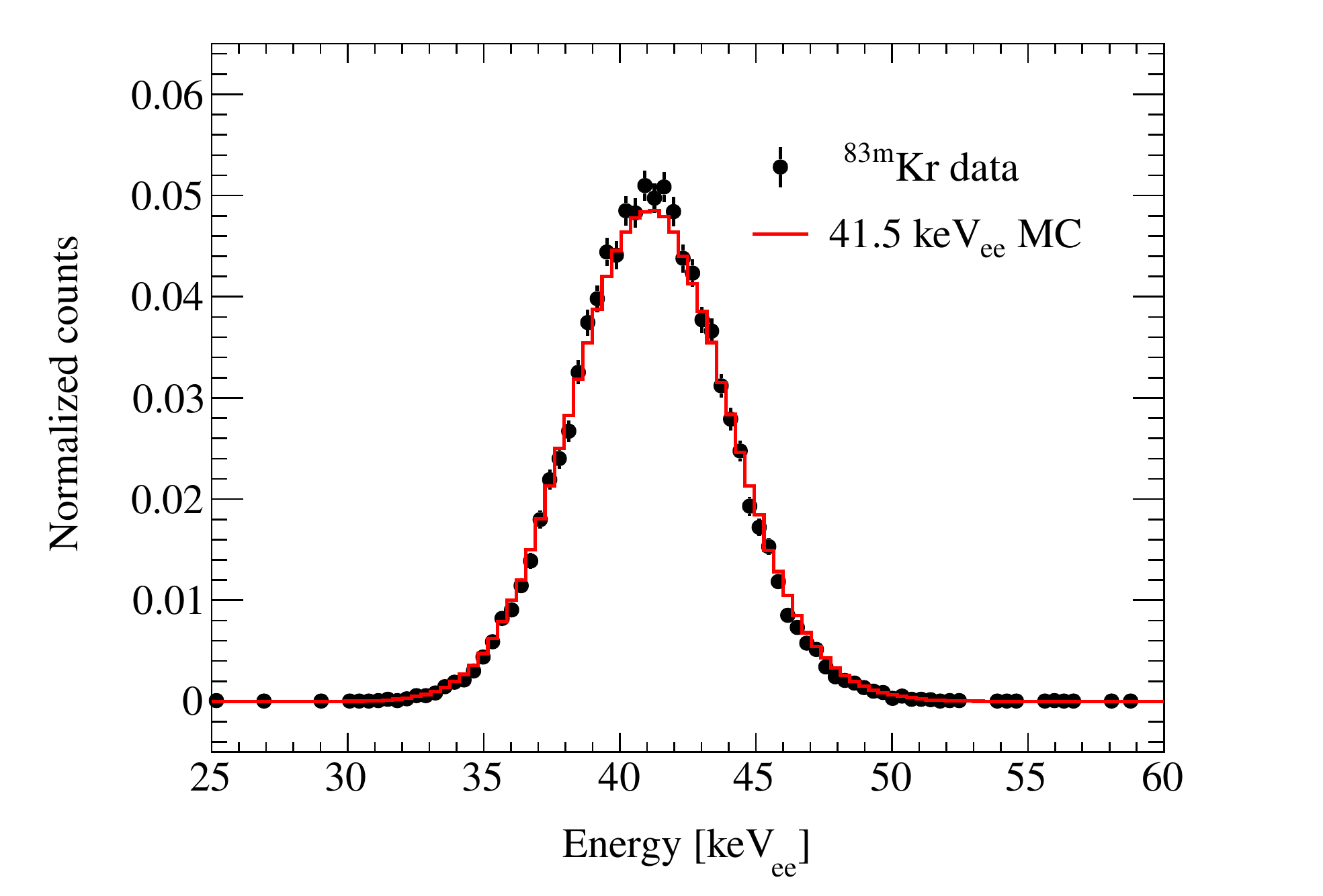}
    \caption{
    Comparison between the ${}^{\rm 83m}$Kr internal conversion electron energy spectrum and the simulation result.
    }
    \label{fig:kr83m_Erec}
\end{figure}

At the end of the PandaX-4T commissioning run, we injected ${}^{\rm 83m}$Kr source (41.5 keV$_{\rm ee}$, electron equivalent energy)~\cite{Zhang2021} into the detector in order to perform a three-dimension uniformity correction for $S1$s and $S2$s. The energy resolution $\sigma_{E}/E$ is $6.8\pm0.1\%$ at 41.5 keV$_{\rm ee}$, in a good agreement with the value ($7.0\%$) given by our ER signal response model, as shown in Fig.~\ref{fig:kr83m_Erec}.

The signal response models in the PandaX-4T are built based on the end-of-run low energy calibrations, including ${}^{220}$Rn, ${}^{241}$Am-Be neutrons, and deuteron-deuteron ($D$-$D$) neutrons~\cite{Meng2021}. The response models follow the construction of the standard NEST~v2.2.1~\cite{NESTv2.2.1, Szydagis2021}. Taking into account all possible detection effects, a simultaneous fit of the ER and NR signal response models is performed, through which the key parameters including the light yield, charge yield, and recombination parameters are determined~\cite{4Tsignalmodel}. 
The PandaX-4T signal response model and the NEST values from a global fit to all available measurements are consistent within the latest reported systematic uncertainties of NEST~\cite{2022APS..APRW09006S,PandaX:2022aac}.

\begin{figure}[htbp]
    \centering 
    \includegraphics[width=0.5\textwidth]{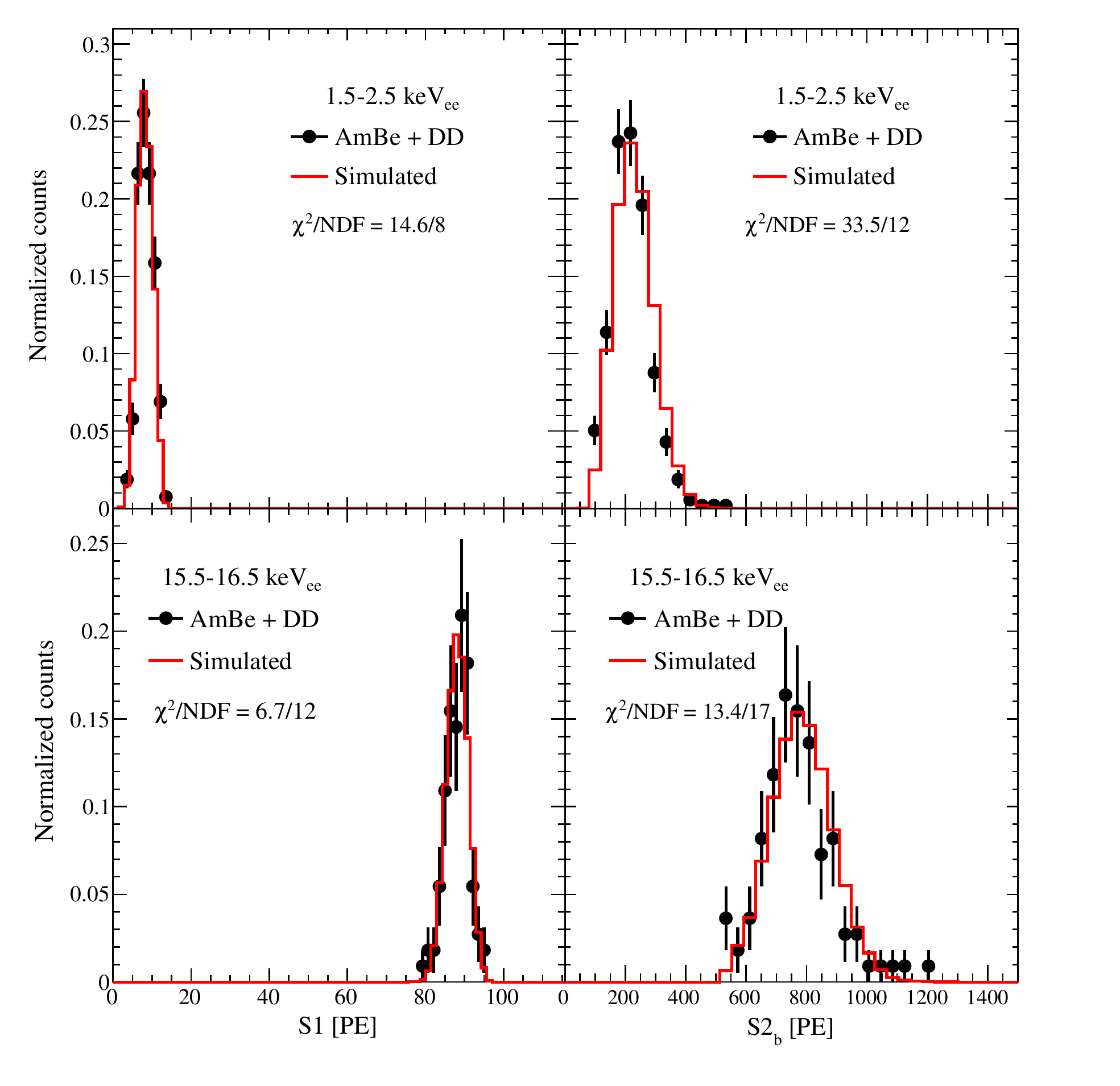}
    \caption{Comparison between the simulated distributions and $^{241}$Am-Be + $D$-$D$ calibration data, using $1.5-2.5$ keV$_{\rm ee}$ (8.2-13.0 keV$_{\rm nr}$) and $15.5-16.5$ keV$_{\rm ee}$ ($68.2-72.1$ keV$_{\rm nr}$) energy windows for illustration. Black dot: normalized $S1$ or $S2_{\rm b}$ distribution of $^{241}$Am-Be and $D$-$D$ calibration data. Red line: simulated distributions.}
    \label{fig:NEST_consistancy} 
\end{figure}

Determining the energy resolution in the region of interest (ROI) is essential for the monopeaked characteristic of the fermionic absorption signal. For the low energy range, due to the lack of a monoenergetic NR calibration source, we compare the distributions of $S1$ and $S2_{\rm b}$ between simulation from our signal response model and the neutron calibration data in a narrow energy window
(scanning from 1 to 16 keV$_{\rm ee}$ with a window size of 1 keV$_{\rm ee}$, which corresponds to the NR energy range from 6 to 70 keV$_{\rm nr}$).
Fig.~\ref{fig:NEST_consistancy} shows such a comparison of NR events in 
two energy windows for illustration.
A good agreement is observed, which indicates that the NR signal simulation from our signal response model is consistent with the data for energy within 70~$\rm keV_{nr}$
The reconstructed energy resolution within ROI given by the simulation can be depicted as
$\sigma_{E_R}/E_{R} = \sqrt{a/E_R + b^2/E_R^2}$,
where $E_R$ is in $\rm keV_{nr}$, $a=0.85~\rm keV_{nr}$ and $b=1.855~\rm keV_{nr}$.
The energy resolution is 49\% (11\%) at 6 keV$_{\rm nr}$ (70 keV$_{\rm nr}$).

\begin{figure}[htbp]
    \centering
    \includegraphics[width=0.48\textwidth]{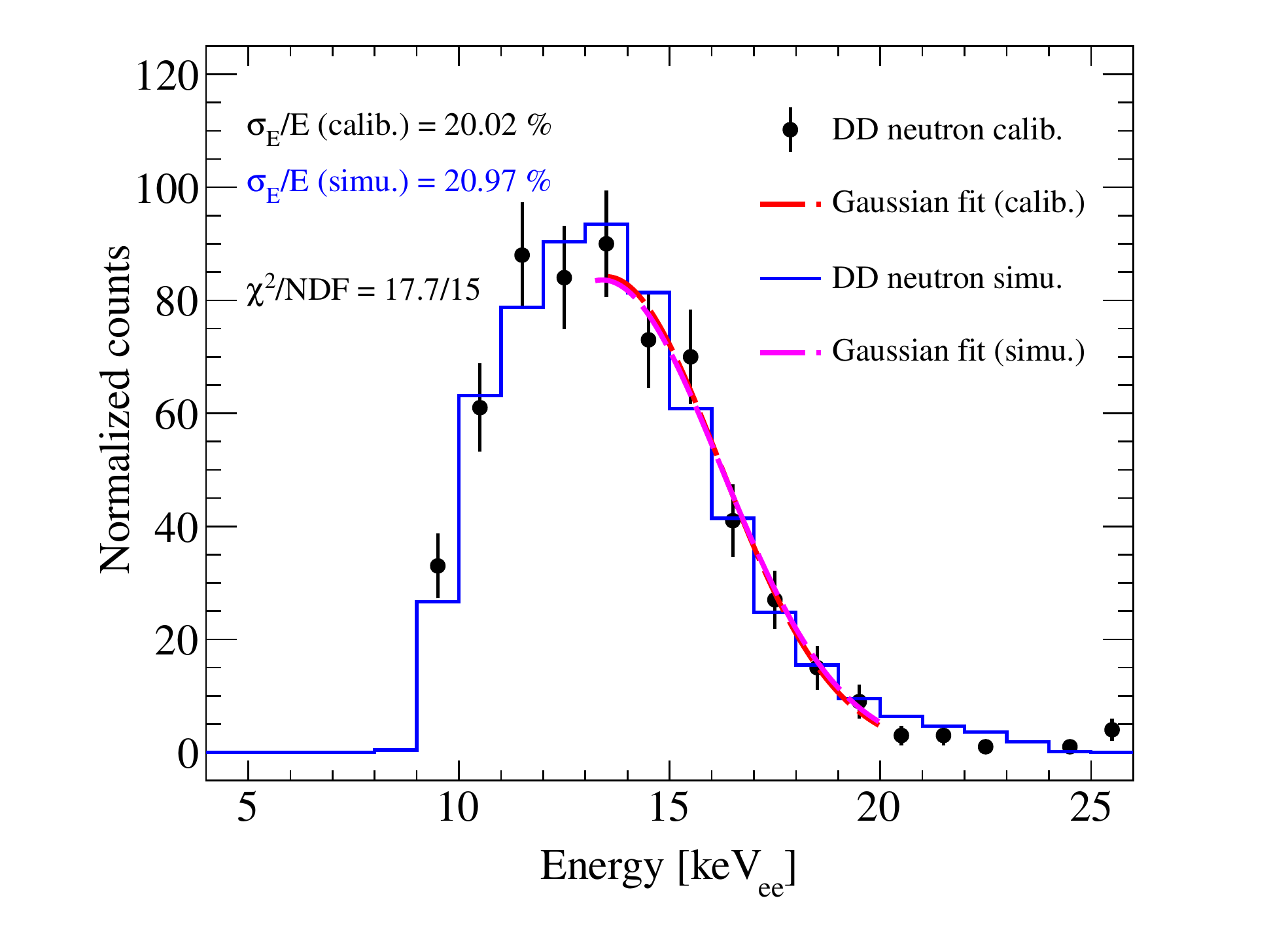}
    \caption{The $D$-$D$ reconstructed energy spectra from calibration and simulation, with a Gaussian fit performing on the right half part of the back-scatter peak (dashed curves). 
    }
    \label{fig:DD-bs}
\end{figure}

The consistency of the NR simulation with the data is further validated through the D-D back-scatter energy peaks.
Simulation neutron events are generated by the PandaX Monte Carlo package BambooMC~\cite{Chen2021} and processed with the signal response model. The reconstructed energy is compared with the data. A Gaussian fit is performed on the right half part of the back-scatter peak, where the detector resolution dominates; see Fig.~\ref{fig:DD-bs}. The fitted width is consistent between the simulation and the data. The relative difference ($\sim5\%$) is taken as a systematic uncertainty of the NR energy resolution.

\begin{figure}[htbp]
    \centering
    \includegraphics[width=0.48\textwidth]{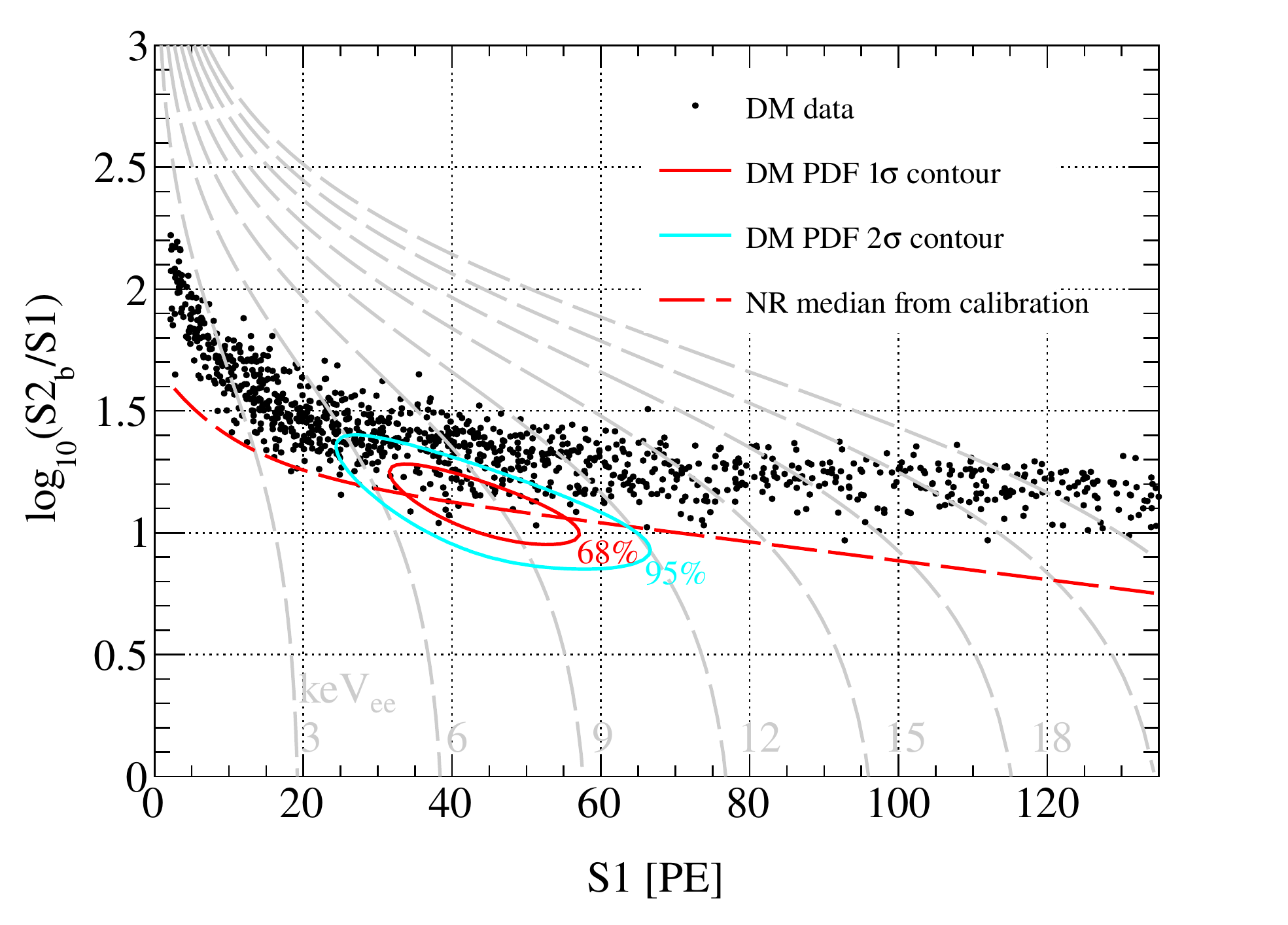}
    \caption{
     Distribution of the selected events in data ROI.
     The red and cyan solid curves illustrate the $68\%$ and $95\%$ contours of the fermionic DM absorption signal distribution with $m_\chi = 100$~MeV/$c^2$. The red dashed line represents the fitted NR median from the end-of-run ${}^{241}$Am-Be and $D$-$D$ calibrations.
    }
    \label{fig:band}
\end{figure}

The physics event selection criteria follow the WIMP search analysis~\cite{Meng2021}, aiming to remove noise, surface backgrounds, accidentally paired events, and events with low-quality signal waveform. In brief, our ROI is selected with $S1$ from 2 to 135 $\rm PE$s, raw $S2$ from 80 to 20,000 $\rm PE$s.
The upper bound of this ROI corresponds to approximately 24 keV$_{\rm ee}$ ($\sim 100$ keV$_{\rm nr}$). 
In total $1058$ events are identified in the ROI from 86.0~live-day exposure data in the PandaX-4T commissioning run, as shown in Fig.~\ref{fig:band}. The background compositions are summarized in Ref.~\cite{Meng2021}, which include tritium, flat ER~(${}^{85}$Kr, Rn, material), surface, ${}^{127}$Xe, neutron, neutrino, and accidental $S1$-$S2$ coincidence events.
The $68\%$ and $95\%$ contours of the probability density function (PDF) for DM mass $m_\chi = 100$~MeV/$c^2$ are overlaid for illustration. The number of observed events within the $68\%$ contour is $26$, and the expected background contribution is estimated to be $21.0\pm2.2$.

The signal of fermionic DM absorption is tested in the ROI, with a two-sided profile likelihood ratio method~\cite{Baxter2021}.
The scanned DM mass parameter ranges from $30$ to $125$~MeV/$c^2$, with the corresponding highest energy deposit peaks at 64 keV$_{\rm nr}$. We construct a standard unbinned likelihood function~\cite{Cui2017, Wang2020} as
\begin{equation}
    \mathcal{L}_{\text{pandax}}=\left[\prod_{n=1}^{n_\mathrm{set}} \mathcal{L}_{n}\right] \times\left[\prod_{b} G(\delta_{b}; \sigma_{b})\right] \times \left[\prod_{p_{*}} G(\delta_{p_{*}}, \sigma_{p_{*}})\right],
\end{equation}
where $n_\mathrm{set} = 5$ as mentioned above, with the single set likelihood function $\mathcal{L}_{n}$ defined below as
\begin{equation}
\begin{aligned}
    \mathcal{L}_{n} =& \operatorname{Poiss}(\mathcal{N}_{\text{obs}}^{n} \mid \mathcal{N}_{\text{fit}}^{n})\\ & \times \left[ \prod_{i=1}^{\mathcal{N}_{\text{obs}}^{n}}\frac{1}{\mathcal{N}_{\text{fit}}^{n}} \Big(N_{s}^{n} P_{s}^{n}(S1^{i}, S2^{i}_{\rm b}\vert\{p_{*}\})\right.\\
    & \left. + \sum_{b} N_{b}^{n}\left(1+\delta_{b}\right) P_{b}^{n}(S1^{i}, S2^{i}_{\rm b}\vert\{p_{*}\}) \Big)\right] .
\end{aligned}
\end{equation}
For each data set $n$, $\mathcal{N}_{\text{obs}}^{n}$ and $\mathcal{N}_{\text{fit}}^{n}$ are the total observed and fitted numbers of events, respectively;
$N_s^n$ and $N_b^n$ represent the amount of DM (signal) and background events; $P_s^n(S1, S2_{\rm b})$ and $P_b^n(S1, S2_{\rm b})$ denote their two-dimensional PDFs. The systematic uncertainties of background estimation ($\sigma_b$) and nuisance parameters ($\sigma_{p_{*}}$) are taken into account via Gaussian penalty function $G(\delta, \sigma)$. 
The systematic uncertainties from NEST on the light yield and charge yield for a low energy NR signal are also considered in the hypothesis tests.
 
\begin{figure}[htbp]
    \centering
    \includegraphics[width=0.48\textwidth]{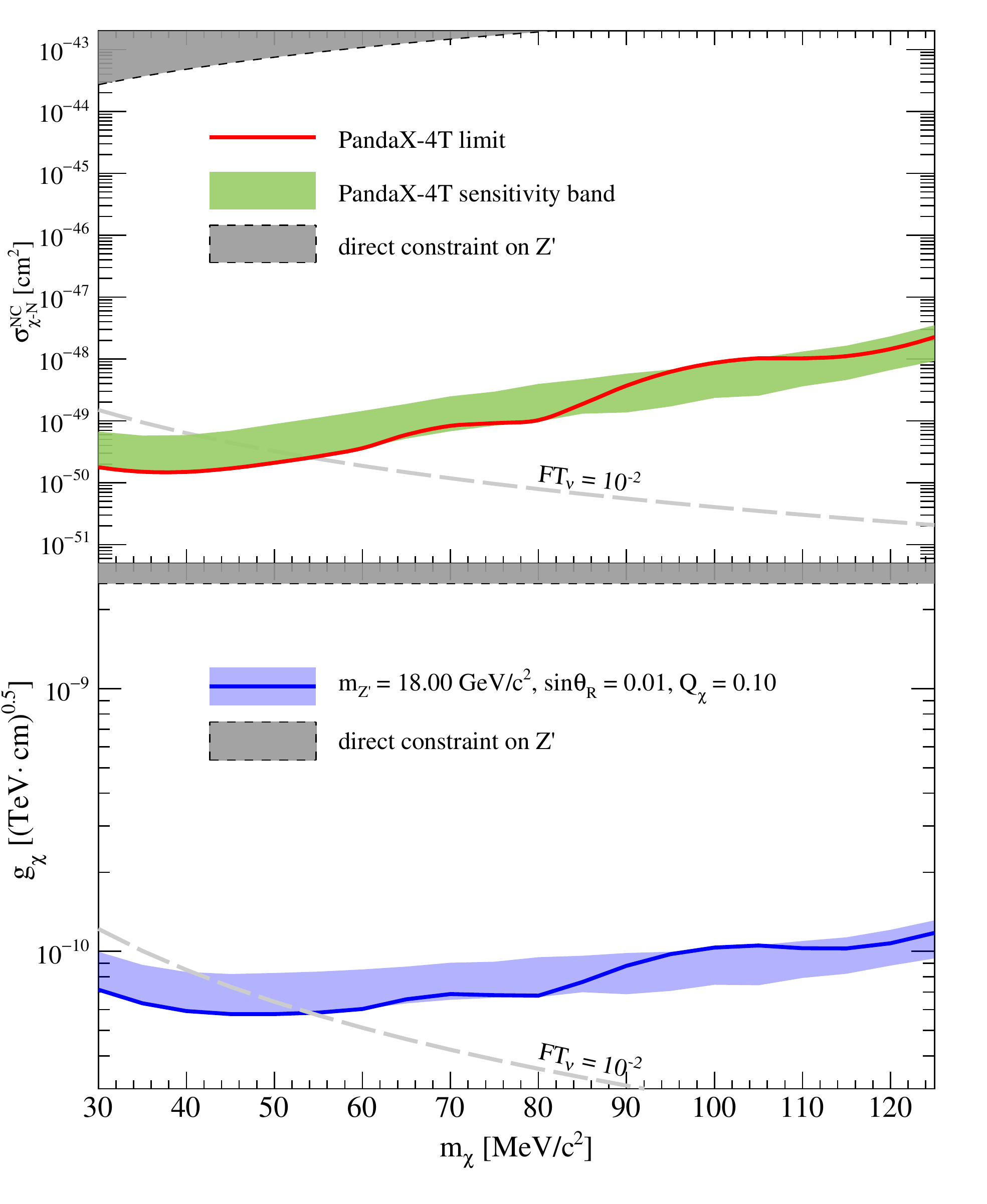}
    \caption{The $90\%$ C.L. upper limits vs $m_\chi$ for the absorption cross sections of fermionic DM from the PandaX-4T commissioning data. The green band represents $\pm 1\sigma$ sensitivity band. The gray shaded area describes the constraints from collider experiment for $Z'$~\cite{Belyaev2019, Dror2019b}. The figure below displays limits and sensitivity bands vs $m_\chi$ for the $U(1)$' gauge coupling from the PandaX-4T commissioning data. The level of fine-tuning needed in the UV complete model to avoid rapid $\chi\rightarrow \nu\nu\nu$ decay is denoted with dashed gray contours labeled $\rm FT_{\nu}$.
    }
    \label{fig:limit_sensBand}
\end{figure}

There is no significant excess identified in the fit. The final $90\%$ confidence level (C.L.) upper limit is shown at the top of Fig.~\ref{fig:limit_sensBand}. This limit curve has a slight downward fluctuation in the DM mass range below $60$~MeV/$c^2$, which is power-constrained to the $-1\sigma$ sensitivity band~\cite{Cowan:2011an}. The strongest limit achieved is $1.5\times10^{-50}$~cm$^2$ at a fermionic DM mass of $40$~MeV/$c^2$.
Direct constraint of $Z'$ from the collider experiments~\cite{Belyaev2019, Dror2019b} is marked by the gray shaded region in Fig.~\ref{fig:limit_sensBand}.

In the UV complete model with a mediator $Z'$, the DM $\chi$ can decay invisibly to neutrinos, $\chi \rightarrow \nu \nu \nu$, which may yield some anomalous change in the equation of the state of the Universe from the era of the cosmic microwave background (CMB) until present day~\cite{Gong2008}. For the interaction parameters adopted in Ref.~\cite{Dror2019b}, namely $m_{Z'} = 18$~GeV/$c^2$, $\sin{\theta_R} = 0.01$, and $Q_\chi = 0.1$, the CMB spectrum given by the WMAP three-year data~\cite{WMAP:2006jqi,WMAP:2006bqn} provides very strong constraints on this model through this triple-neutrino decay channel. To avoid this indirect detection bound, fine-tuning is needed~\cite{Dror2019b}; the corresponding level is shown in Fig.~\ref{fig:limit_sensBand}.
Given the relationship between $U(1)$' gauge coupling $g_\chi$ and the energy cut-off scale $\Lambda$,
 the obtained upper limits of cross-section in the top of Fig.~\ref{fig:limit_sensBand} can be directly translated into the constraints on the coupling $g_\chi$, with an order of $10^{-10}$~$(\rm TeV\cdot cm)^{1/2}$, as shown in the bottom of  Fig.~\ref{fig:limit_sensBand}. 


In summary, we explore the neutral current absorption signals of fermionic DM in the PandaX-4T 0.63-tonne-year exposure data in its commissioning run, which is the first search for a monoenergetic NR signature performed in direct detection experiments. No significant excess is observed above the expected background. A new model-independent exclusion limit is set on the sub-GeV DM-nucleon interactions, excluding the scattering cross section with nucleon as low as $1.5\times10^{-50}$~cm$^2$ for DM mass of $40$~MeV/$c^2$.
Together with cosmology indirect detection and collider search, this result provides strong constraints on the UV complete model with a $Z'$ mediator.
Searching for light fermionic DM absorption interaction with electrons is also performed~\cite{PandaX:2022ood}. PandaX-4T continues taking more physics data and is expected to improve the sensitivity by another order of magnitude with a 6-tonne-year exposure.



This project is supported in part by grants from the National Science Foundation of China (Grants No. 12090060, No. 12090061, No. 12005131, No. 11905128, No. 11925502, No. 11775141), a grant from the Ministry of Science and Technology of China (Grant No. 2016YFA0400301), and by Office of Science and Technology, Shanghai Municipal Government (Grant No. 18JC1410200). This project is also funded by China Postdoctoral Science Foundation (No. 2021M702148). We thank the Double First Class Plan of the Shanghai Jiao Tong University for support. We also thank the sponsorship from the Chinese Academy of Sciences Center for Excellence in Particle Physics (CCEPP), Hongwen Foundation in Hong Kong, Tencent Foundation in China and Yangyang Development Fund. Finally, we thank the CJPL administration and the Yalong River Hydropower Development Company Ltd. for indispensable logistical support and other help.

\bibliography{apssamp}

\end{document}